\documentclass[prl,showpacs,twocolumn]{revtex4-2}

\usepackage{epsfig}
\usepackage{color}
\usepackage{graphicx}
\usepackage{amssymb,amsfonts,amsmath}
\begin{document}
\title{Direct observations of causal links in plastic events validates statistical analysis tools for seismology}

\author {Pinaki Kumar$^1$, Roberto Benzi$^2$, Jeannot Trampert$^3$ and Federico Toschi$^{1,4}$}

\affiliation{$^{1}$Department of Physics, Eindhoven University of Technology, \\PO Box 513, 5600  MB, Eindhoven, The Netherlands and CNR-IAC, Rome, Italy.}
\affiliation{$^{2}$Department of Physics, University of ''Tor Vergata``, \\Via della Ricerca Scientifica 1, 00133 Rome, Italy.}
\affiliation{$^{3}$Department of Earth Sciences, Utrecht University, \\ PO Box 80115, NL-3508 TC, Utrecht,The Netherlands.}
\affiliation{$^{4}$CNR-IAC, 00185, Rome, Italy.}

\date{\today}
\pacs{47.27.-i, 47.27.te}

\begin{abstract}
Earthquakes are complex physical processes driven by the stick-slip motion of a sliding fault. After the main quake, a series of aftershocks typically follows. These are loosely defined as events that follow a given event and occur within prescribed space-time windows. In seismology, it is however impossible to establish a causal relation and the popular Nearest-Neighbor metric is commonly used to distinguish aftershocks from independent events. Here we employ a model for earthquake dynamics, previously shown to be able to correctly reproduce the phenomenology of earthquakes, and a technique that allows us to separate independent and triggered events. We show that aftershocks in our catalogue follow Omori’s law and we employ the model to  show that the Nearest-Neighbor metric is effective in separating independent events from aftershocks.
\end{abstract}

\maketitle

\section{Introduction}
Aftershocks are earthquakes that follow a given event and occur within prescribed space-time windows. The notion of aftershocks, therefore, implies causality. Separate fields of science dealing with dynamical systems treat causality very differently,  varying from a theoretical notion in physics \cite{dariano2018} to a common-sense approach in the legal domain \cite{summers2018common},  via complex system analysis in epidemiology \cite{galea2010causal} to name just a few.
Establishing causality between events can take the form of direct perception, logical reasoning, or statistical tests of varying sophistication. In certain cases, after careful exclusion of confounding variables, one observes that when a certain action/event occurs, another event always occurs later in time and when the former event is missing, the latter never occurs. This can be labelled as direct evidence of causality. When elimination of confounding variables is not easy or direct observation of the dynamic processes is not possible, one is often forced to resort to statistical testing which necessarily involves ad-hoc thresholds, above which causal claims can be made. In this paper, we will present direct observations of causal links between plastic events in a soft-glass, which appear to be a good proxy for earthquakes \cite{benzi2016earthquake}, and by analyzing the statistics of our unambiguous causal catalogue, we provide a validation to widely used statistical tools in seismology.

Earthquakes cluster in the space-time-magnitude domain, which is most readily observed by sequences of aftershocks said to be triggered by a prior main event. The rate of occurrence of aftershocks follows the well-known Omori law \cite{utsu1995centenary}. The causal link implied between mainshock and aftershock is assumed to be through static or dynamic triggering \cite{freed2005},  but the direct observation of this link via relevant dynamic variables inside the Earth remains elusive. The techniques to identify aftershocks have included space-time window based approaches \cite{gardner1974seq,Petersen2016,knopoff1982b}, stochastic declustering of earthquakes modeled as a point process \cite{zhuang2004analyzing}, evolving random graphs \cite{abe2004scale}, machine learning techniques like diffusion maps \cite{bregman2019aftershock} and neural networks \cite{zhang2019aftershock} and a Nearest-Neighbor (NN) rescaled space-time-magnitude metric \cite{baiesi2004scale, zaliapin2008}. This last method, which we will apply to our aftershock database, has seen wider adoption in seismology \cite{gu2013triggering, zaliapin2013,zhang2016new,moradpour2014nontrivial,benzi2016earthquake,zaliapin2020}.  The NN metric combines the phenomenological Gutenberg-Richter (GR) law with the fact that earthquake epicenters have a fractal distribution and is an estimate of the number of expected events within a certain radius, time and magnitude range \cite{baiesi2004scale}. The probability distribution of this NN measure is close to bi-modal, with one mode corresponding to that of a Poissonian field and the other to clustered events.

Previously, we inferred that Omori's law, governing aftershock occurrence, held true for our system \cite{benzi2016earthquake} by accepting that the assumptions underlying the NN-measure used in seismology were applicable to our numerical system as well. The identification of causal relations between plastic events in our previous work was thus statistical. In this paper, our aim is to go in the opposite direction - we first find a means of establishing causality by direct observation, then we check whether aftershocks in our system indeed follow Omori's law, and finally whether the assumptions behind the NN measure are valid. 

\section{Method}
\begin{figure} 
\centering
\includegraphics[width=0.47\textwidth]{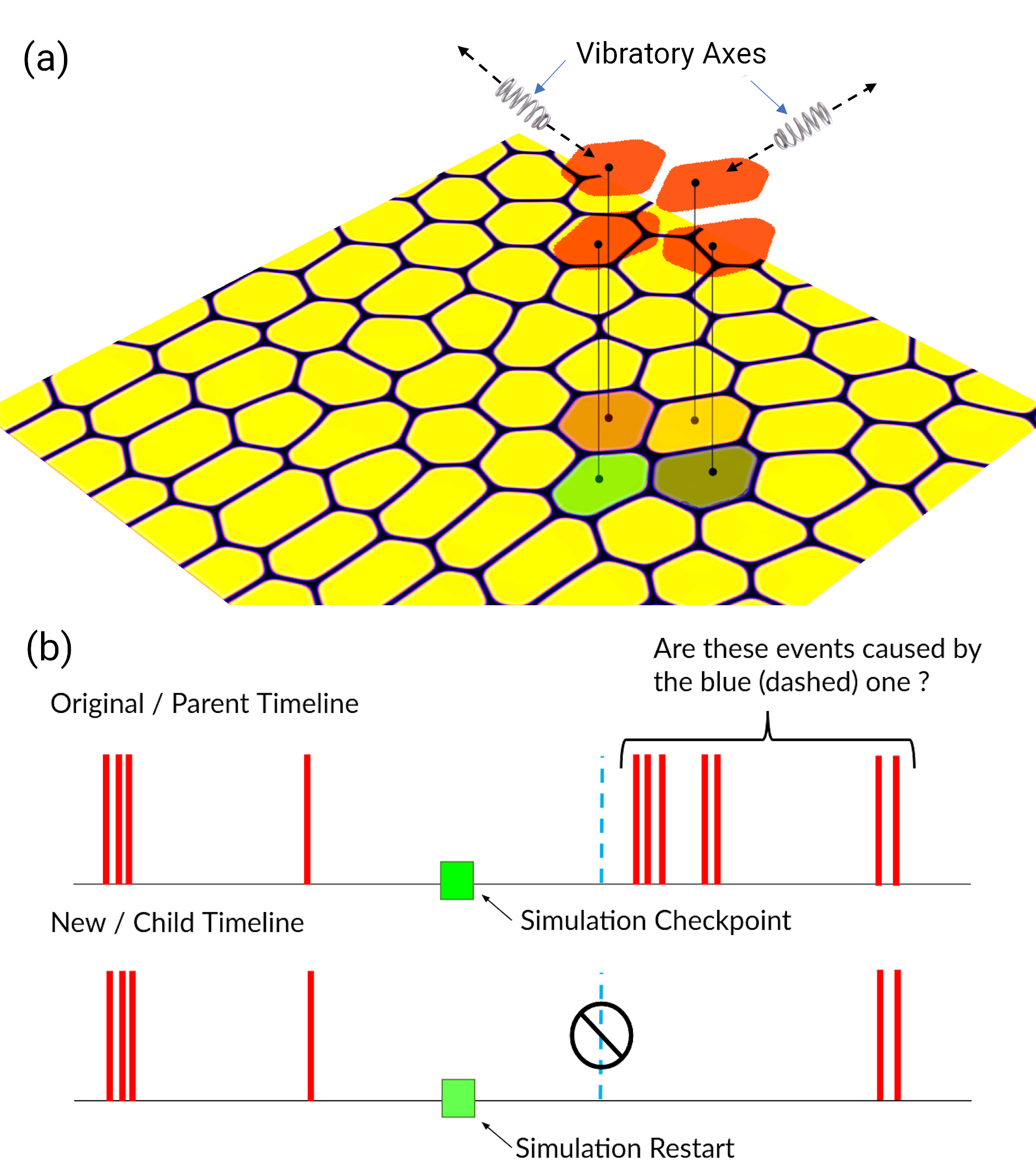}
\caption{(a) Density field of our binary fluid soft-glass with drops of one component dispersed at a high packing fraction in the other. The geometrical shape of a quartet of neighboring drops about to undergo a topological rearrangement is extracted. Each drop is subjected to a vibratory forcing along an axis from the drop's center to the quartet's center of mass thus turning the local potential energy hill into a valley and arresting the droplets. (b) Schematic approach for establishing causality. A simulation labelled the ``parent'' is run and events are identified. To find the causal aftershocks of a given event, a ``child'' simulation where this event is arrested is run starting from a checkpoint of the LB population just prior to said event. All subsequent events, which are present in the parent timeline but missing from the child timeline are labelled aftershocks or if otherwise, independent. Figure reproduced from author's own work \cite{kumar2020multi}.}
\label{fig-scheme}
\end{figure}

To investigate event causality, we adopt a numerical scheme based on a multi-component Lattice Boltzmann (LB) method \cite {benzi2009mesoscopic}, which we have shown to be a good in-silico earthquake proxy \cite{benzi2016earthquake}. We briefly describe the method here. Our system comprises a binary fluid, with competing long and short range inter- and intra-fluid interactions which promote frustration of the interface. At higher packing fractions, glassy dynamics sets in with very long relaxation times resulting in a soft-glass (see \figurename{ \ref{fig-scheme}} (a)), ageing behavior \cite {benzi2009mesoscopic} and Herschel-Bulkley rheology \cite{benzi2010herschel}. The system relaxes through irreversible topological rearrangements of neighboring droplets which deform plastically. These plastic events, always comprising 4 droplets (a quartet) in 2D, radiate a part of the released energy away from the site and are accompanied by stress drops \cite{benzi2014direct}. When experiencing a Couette shear slightly below the yield stress, the system exhibits a long-range correlation in stress and the plastic event sizes, duration and inter-event times follow well established empirical seismic scaling laws \cite{benzi2016earthquake, kumar2020interevent}. 

Our method for establishing direct causality is sketched in \figurename{ \ref{fig-scheme}} (b). We first run our LB simulation with regular checkpoints of the LB populations and collect plastic event data, and this simulation run is labelled the ``parent-run''. For each event identified in the parent, we re-run another LB simulation called the ``child-run'' where the simulation is restarted from a checkpoint just prior a target event, and where that event is stopped from occuring. We then check to see how the downstream dynamics differ in the child run - specifically which events from the parent-run disappear and which remain the same.

In \cite{kumar2020multi} we presented a method to stop a plastic event from occurring inspired by the stabilization of the Kapitza pendulum. Kapitza \cite{kapitza1965dynamical} showed that fast vibrations with small amplitude could stabilize an inverted pendulum by separating the resulting vibrations into slow and fast components. Here the term fast is relative to the natural frequency of the system which is slower. Kapitza then showed that the faster vibrations effectively modified the local potential landscape so that the vertical position became a point of stable equilibrium. In a similar vein, we applied small fast vibrations to the quartet of drops about to undergo a plastic event. This lowered the local potential energy hill, and since events only occur as a way to release energy, this prevented the event from occurring. This topological arrest of a quartet required modifications to the usual LB scheme as : 

\begin{equation}\label{eq1}
f^{*} = (1-\epsilon)\underbrace{\left(f - \frac{1}{\tau}(f - f^{\mathrm{eq}}) + F\right)}_{\text{Normal LB update}} + \epsilon(f^{\mathrm{eq-Arrest}})
\end{equation}

The normal LB update is highlighted, with $f$ representing the LB population distribution, $\tau$ being the time to relax to the local Maxwellian equilibrium $f^{\mathrm{eq}}$ and $F$ is the intra- and inter-molecular forcing term. We modify this and introduce the term $f^{\mathrm{eq-Arrest}}$, which corresponds to the equilibrium population distributions of the set of points making up the droplet quartet immediately prior to the event, and a scalar $\epsilon$ which is a spatio-temporal function - spatially it takes the value of unity at the points belonging to the drop quartet and zero elsewhere while the temporal part is periodic with magnitude between 0 and 1. In effect what is achieved is an operation where the quartet is frozen in time and space i.e. arrested and then released periodically. This can be visualized in \figurename{ \ref{fig-scheme}} (a) by imagining the four drops to be vibrating along an axis connecting their individual center of mass and the collective center of mass of the quartet. The amplitude of the vibrations is very small and is of the order of the random diffusive current in the interface, and we have demonstrated in \cite{kumar2020multi} that this technique does not disturb neighboring drops and allows us to achieve a ``clean'' arrest of an event.

Using the modified LB scheme, we built a database of event sequences with the aim to analyze a large number of these sequences to extract causal relations. In this database, each event is associated to two simulations - a \textit{parent} run where the event occurs and a \textit{child} run where the event has been prevented from occurring. The criteria for classifying event pairs as being either causally connected or independent is as follows: those events from the \textit{parent} timeline which do not occur in the \textit{child} timeline subsequent to a plastic event having been stopped, are said to be causally connected or in the language of seismology are aftershocks. On the other hand, those events from the \textit{parent} timeline which re-occur at the exact same time and location in the \textit{child} timeline despite the plastic event being stopped, are said to be independent events. The premise behind the above classification is simple. A change in the local topology due to a parent event manifests itself as a propagating disturbance and becomes the cause for triggering those events which have now disappeared after the parent has been stopped. In a similar vein, independent events have their origin in other factors but not the stopped event. All event pairs in our database are classified accordingly, yielding over 1300 event-aftershock pairs. When a single event has multiple aftershocks we call it an aftershock sequence.
Some sequences have several events and the causal chain can span a long time duration (up to 400k timesteps), while in other cases the chain terminates more quickly (30k timesteps).

\section{Results}
\begin{figure} 
\centering
\includegraphics[width=0.47\textwidth]{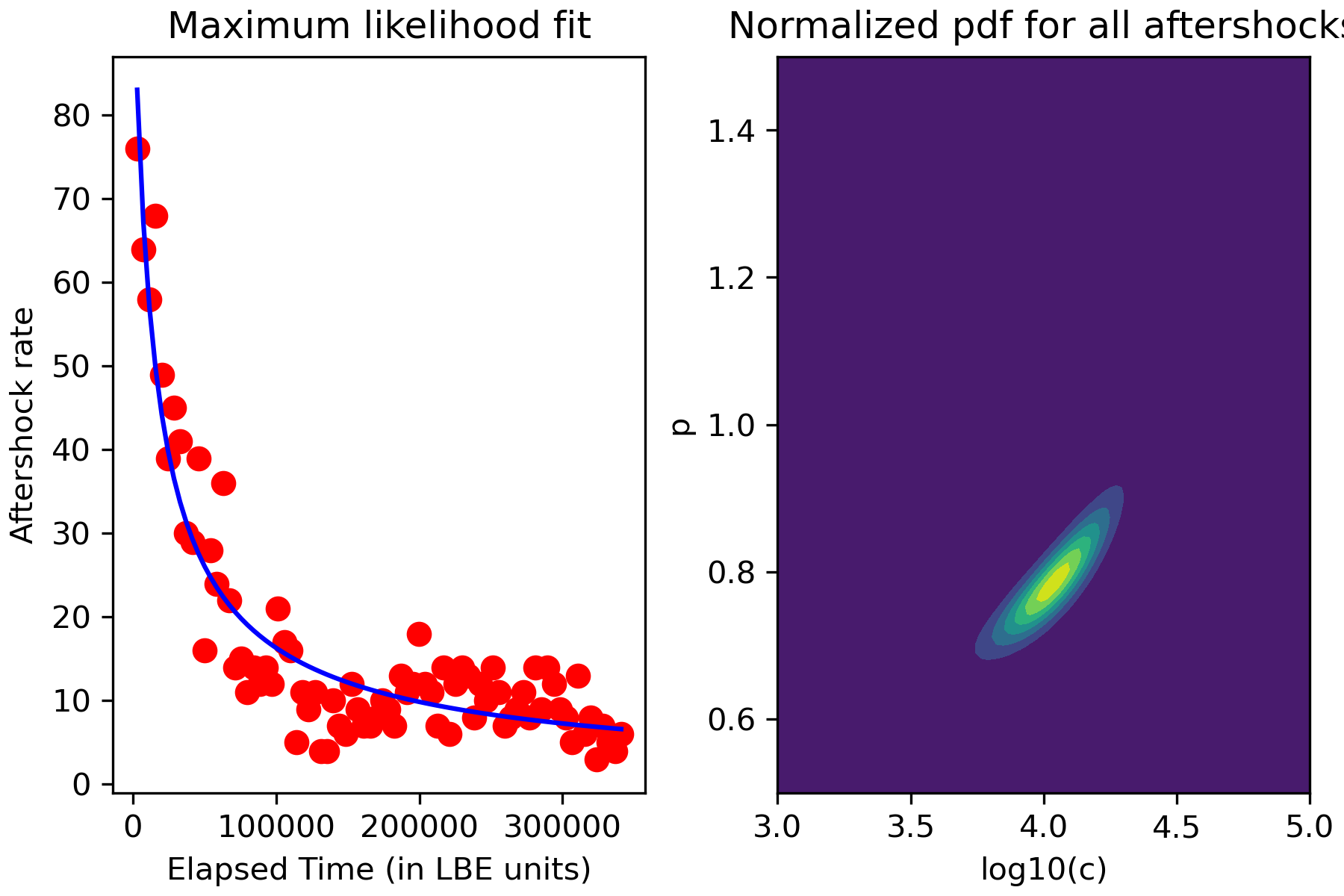}
\caption{Left panel: Aftershock rate against the time elapsed since the parent event as red circles in the complete catalogue without a preset maximum elapsed time. The blue line corresponds to the curve from the Bayesian maximum likelihood parameters. Right panel: Normalized joint probability density for $p$ and $c$ (from Eq.\eqref{eq3}) for all aftershocks in our database.  The yellow contour has a value of 1 and the purple one a value of 0.}
\label{fig-omori}
\end{figure}

For every sequence or causal chain in our database, we calculated the time elapsed between the \textit{parent} event and its \textit{child} (aftershock) and this time difference is used to fit Omori's law. Omori's law, originating in seismology, as shown in Eq.\eqref{eq3}, states that the rate $r$ of aftershocks resulting from a parent earthquake scales roughly with the inverse of time. In practice, from earthquake catalogs it has been estimated that the exponent $p \in [0.8,1.5]$ with productivity $k$ and time delay $c$ being constant for a given system \cite{utsu1995centenary}. 

\begin{equation}\label{eq3}
r = \frac{k}{(c+t)^{p}}
\end{equation}

where $t$ is the time elapsed since the \textit{parent} event.  Given that we have access to the precise occurrence times of events and their aftershocks, evaluating the aftershock rate is straightforward. Inferring the 3 parameters in Omori's law, however, is an ill-posed optimization problem full of trade-offs.  We opted for a Bayesian estimation \cite{holschneider2012}, including a reparametrization, which separates the determination of the productivity $k$ from the shape parameters $p$ and $c$. Furthermore in this reformulation, the shortest and longest elapsed times between parent and aftershock in the catalogue determine a large part of the uncertainty in the shape parameters \cite{holschneider2012}
In \figurename{ \ref{fig-omori}}, we plot the aftershock rate versus time elapsed since the \textit{parent} event. Our Bayesian inference gives $p=0.78 \pm 0.03$, $c=(10 \pm 4)10^3$ and $k=(145 \pm 32)10^3$.  There is a trade-off between $p$ and $c$, but they do not depend on the maximum elapsed time used in the analysis. $k$ is independent of the shape parameters and increases linearly with the number of events in catalogue determined by the maximum elapsed time \cite{holschneider2012}.

\begin{figure} 
\centering
\includegraphics[width=0.47\textwidth]{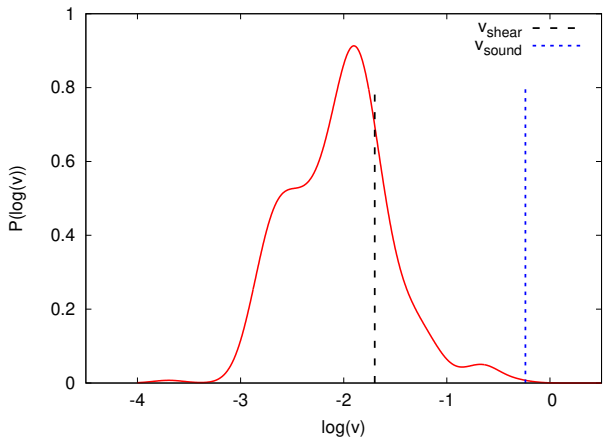}
\caption{Probability distribution of the log of speed ($v$) of information propagation to the first aftershock. A kernel density estimate is applied to obtain the smooth distribution shown. The black dashed line indicates the speed of propagation of elastic shear waves ($v_{shear }= 0.02$) and the blue dotted line shows the speed of sound ($v_{sound} = 0.58$) in the lattice-Boltzmann scheme.}
\label{fig-speed}
\end{figure}

Since we know the position and timing of the aftershocks with respect to the main event, we can analyze the speed of information propagation $v$ between the pairs. In this analysis, we only consider the first aftershock to each parent to avoid secondary effects due to triggering from other aftershocks and wave reflections from domain walls, which could bias estimates of information speed.  In \figurename{ \ref{fig-speed}}, we present the probability distribution of $\log v$. Also shown are two speeds - $v_{shear}$ and $v_{sound}$ representative of our system.
Our implementation gives us access to the bulk thermodynamic pressure and under the constraints of mass and momentum conservation and lattice isotropy, the speed of sound is then simply $v_{sound}=\sqrt{dP/d\rho}=1/\sqrt{3}$. Since we are forcing our soft-glass in a regime just below the yield stress, it has the properties of a solid, and thus can sustain both compression and shear waves, unlike fluids that cannot transmit shear waves. 
$ v_{shear} \approx 0.02$ has been determined previously for our system \cite{benzi2013}.
Roughly 80\% of first aftershocks fall within the light-cone of the elastic shear wave while all first aftershocks fall within the light-cone of the speed of sound. The fact that almost 20\% of events are triggered before the elastic shear waves can reach them is interesting. On closer inspection, we found that for those events,  the involved droplets are either directly in contact with the droplets belonging to the parent event or are 2nd neighbors. Due to the constraints of geometry imposed by the rigid interface these neighbor drops are forced to deform almost simultaneously with the parent event compared to drops that are farther away.  In summary, all aftershocks in our system are triggered by two means - (i)  by geometrical constraints (of a rigid interface) arising from the deformation of neighbours who are undergoing a plastic event and (ii) passing elastic shear wave perturbations that overcome the local balance of forces.

\section{Discussion}
\begin{figure} 
\centering
\includegraphics[width=0.47\textwidth]{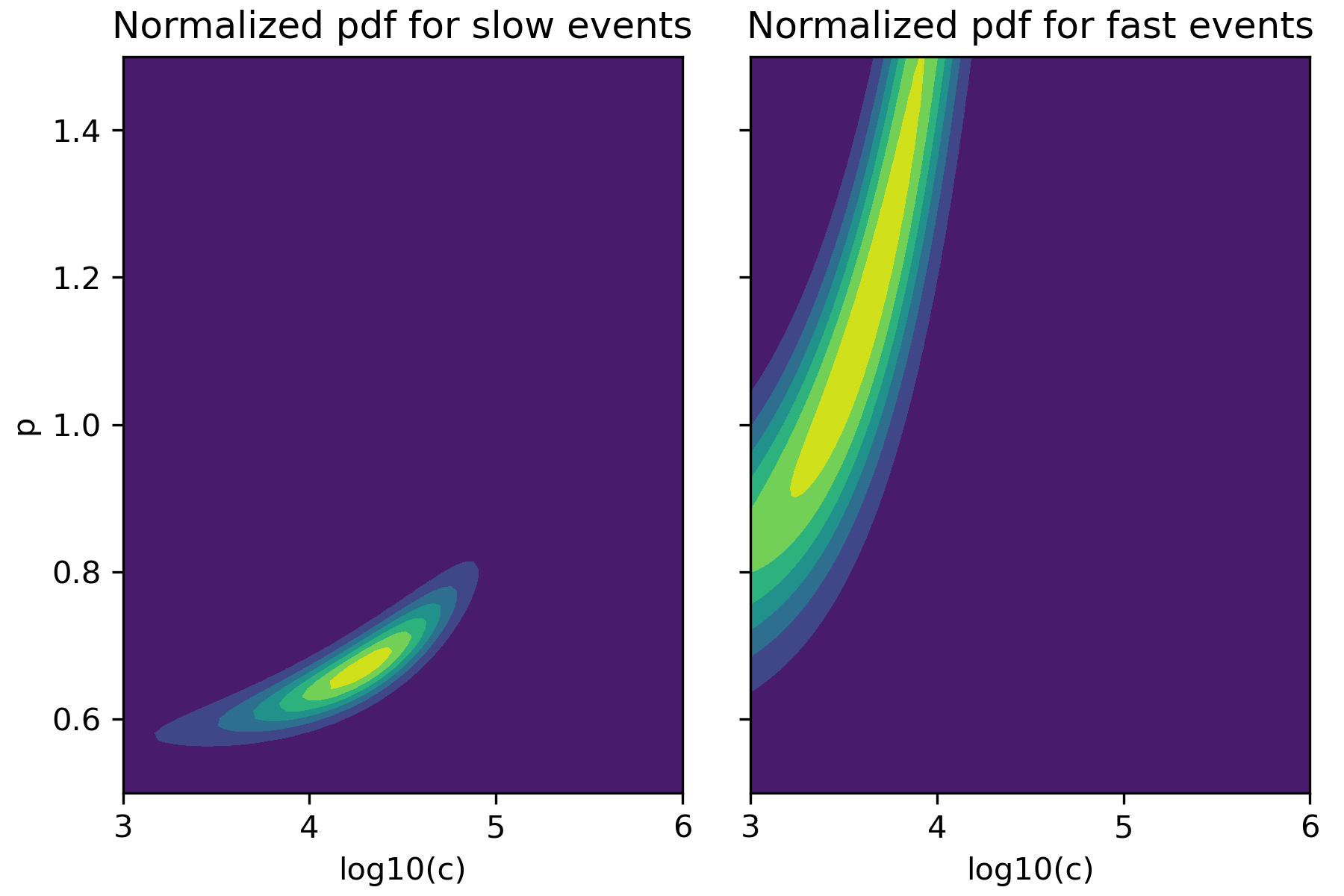}
\caption{Normalized joined probablity density for $p$ and $c$ (from Eq.\eqref{eq3}) for events triggered  by contact (fast) mode and shear wave (slow) mode.  The yellow contour has a value of 1 and the purple one a value of 0.}
\label{fig-omori2}
\end{figure}

We have identified two modes of event triggering, by contact and by passage of shear waves. This is similar to (quasi-)static versus dynamic stress triggering identified in seismology \cite{hill2015,freed2005}.  Static stress triggering refers to the stress change just before an earthquake and after the dynamic stresses transported by seismic waves have dissipated, which then triggers an event on a nearby fault close to the Coulomb failure threshold. Quasi-static triggering takes viscous dissipation of static stresses into account.  The oscillatory dynamic stresses transported by elastic waves,  have a longer reach, and especially surface waves can bring a region close to failure.  We observe that events triggered by seismic waves can effect the whole medium and hence are spread out in time, while the contact events (static stress changes) act locally and fast (\figurename{ \ref{fig-speed}}). Having clearly identified the two classes of triggered events, we fitted Omori's law to them both independently (\figurename{ \ref{fig-omori2}}). The interesting observation, which is worthwhile investigating in seismology, is that shear wave triggered  events have a notably lower Omori slope $p=0.67 \pm 0.03$ (event rate decays more slowly) than contact events $p=1.14 \pm 0.10$. The time delay $c$ and the event productivity $k$ are statistically not separable for the two modes.

\begin{figure} 
\centering
\includegraphics[width=0.47\textwidth]{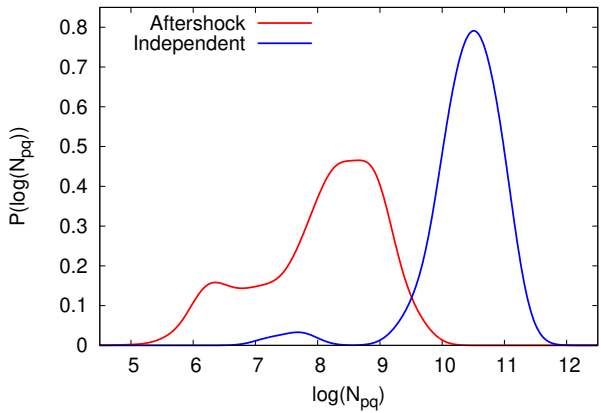}
\caption{Probability distribution of the nearest-neighbor ``distance'' measure, $N_{pq}$ of Eq. \eqref{eq4} with the red curve representing aftershocks and the blue curve representing independent events. The major peaks for the two scenarios show a clear bi-modality of the combined distribution, as in the case of earthquakes. The highest peak of the aftershock distribution corresponds to shear wave triggered events and the smaller peak to contact aftershocks.}
\label{fig-zpin}
\end{figure}

Aftershock identification in the case of earthquakes is more subjective and mostly statistical.  Having a ground truth aftershock catalogue,  puts us in a position to evaluate the popularly used statistical approach of the Nearest-Neighbor (NN) metric \cite{zaliapin2008} in seismology.  For every pair of events $(i,j)$ in the event database, such that $i$ occurred before $j$, the NN  measure is defined as :

\begin{equation}\label{eq4}
N_{ij} = \Delta T ( \Delta L )^{d_{f}} 10^{-B_{GR} log M}
\end{equation}

where $\Delta T = T_{j} - T_{i}$ is the inter-event time, $\Delta L$ is the spatial separation, $d_{f}$ is the fractal dimension of event centers, $B_{GR}$ is the slope of the Gutenberg-Richter law and $M$ is the magnitude of the event. All of these quantities are directly available from our database of events. Through empirical observations \cite{zaliapin2013}, it was shown that the distribution of $N_{ij}$ for real earthquakes exhibits a bi-modality,  which permits to discriminate between aftershocks and independent events. Since, we already know which events are aftershocks and which are independent, we plot the respective NN metric distribution in \figurename{ \ref{fig-zpin}}. The distributions of $N_{ij}$ for aftershocks and independent events mainly show two distinct peaks. Thus we can state that the bi-modality that emerges in the distribution of all event pairs is not a statistical artifact but emerges naturally from the combination of the distributions of aftershocks and independent events. Note that a synthetic earthquake catalog generated numerically from a Poisson point process would give a uni-modal distribution.

\section{Conclusion}
We devised a technique, which allows us to separate plastic events into independent and triggered events or aftershocks.  A Bayesian analysis showed that the aftershocks in our catalogue follow Omori's law.  We further noticed that aftershocks are triggered by two different modes: by contact and by a passing shear wave. While the two classes of aftershocks independently follow Omori's law, their temporal decay is significantly different. We suggest that this might allow seismologists to separate dynamically from statically triggered aftershocks as well. Finally, we showed that,  in the absence of a direct way to establish causality, the popular Nearest-Neighbor metric used in seismology is indeed a good way to separate aftershocks from independent events.

\bibliography{mainbib}

\end{document}